\documentclass[12pt]{iopart}

\usepackage{epsfig}

\begin{document}

\title[Li $et al$., Effect of Zn impurity in LaFeAsO$_{1-x}$F$_x$]{Effect of a Zn impurity on $T_c$ and its implication to pairing symmetry in
 LaFeAsO$_{1-x}$F$_x$}

\author{Yuke Li$^1$, Jun Tong$^{1}$, Qian Tao$^{1}$, Chunmu Feng$^{1}$, Guanghan Cao$^{1,2}$, Weiqiang Chen$^3$, Fu-chun Zhang$^3$, Zhu-an Xu$^{1,2}$\footnote[1]{Electronic address: zhuan@zju.edu.cn}}

\address{$^1$ Department of Physics, Zhejiang University,
Hangzhou 310027, China}
\address{$^2$ State Key Laboratory of
Silicon Materials, Zhejiang University, Hangzhou 310027, China}
\address{$^3$ Department of Physics and Center
of Theoretical and Computational Physics, The University of Hong
Kong, Hong Kong, China}

\date{\today}

\begin{abstract}
The effect of non-magnetic Zn impurity on superconductivity in
LaFe$_{1-y}$Zn$_y$AsO$_{1-x}$F$_x$ system is studied
systematically. In the presence of Zn impurity, the
superconducting transition temperature increases in the
under-doped regime, remains unchanged in the optimally doped
regime, and is severely suppressed in the over-doped regime. Our
results suggest a switch of the symmetry of the superconducting
order parameters from a $s$-wave to $s_{\pm}$ or $d$-wave states
as the charge carrier doping increases in FeAs-based
superconductors.
\end{abstract}


\maketitle
\section{Introduction}
Discovery of iron-based superconductivity \cite{Kamihara08} has
generated great interest in condensed-matter physics. Similar to
the high transition temperature superconducting (SC) copper
oxides, the parent compounds of the iron pnictide are
antiferromagnet. Superconductivity emerges upon chemical doping,
which introduces charge carriers and suppresses the
antiferromagnetic (AFM) order. It is generally agreed that
magnetism plays an important role in the superconductivity of both
cuprates and pnictides.  On the other hand, iron-based
superconductors also show different behaviors from the cuprates.
One of their major differences is the pairing symmetry. It has
been well established that the SC pairing in cuprate is of nodal
$d$-wave symmetry\cite{Tsuei1,Wollman}. In pnicitdes, the pairing
symmetry continues to be an important and outstanding issue. There
are clear experimental evidences for full SC gaps in FeAs-based
superconductors\cite{Andreev,ARPES}. Within the full gap scenario,
because of the multi Fermi surfaces\cite{Singh}, the relative
phases of the SC order parameters in hole or electron pockets can
be either positive ($s$-wave pairing) or negative
($s_{\pm}$-pairing)\cite{Mazin,Kuroki,JPHu,Tes,LeeDH,ChenWQ},
depending on the sign of the inter-Fermi pocket pair scattering
amplitude or their Josephson coupling. The $s_{\pm}$ pairing is
appealing with some experimental supports \cite{Tsuei2}. In
addition, there are also evidences for $d$-wave SC gap in
FeP-based superconductors\cite{Moler}.

Non-magnetic impurity is an important probe to pairing symmetry.
Non-magnetic impurities do not cause severe pair-breaking effect
in a conventional $s$-wave superconductor according to the
Anderson's theorem\cite{Anderson}. In the $s_{\pm}$ state, the
order parameters in hole and electron pockets have opposite signs,
non-magnetic impurities like Zn can severely suppress the SC
transition temperature $T_c$, similar to the effect in high-$T_c$
cuprates with $d$-wave pairing\cite{Bang,Kontani}.

Zinc element has a stable $d^{10}$ configuration in the alloy
\cite{Singh2009}, and can serve as the best non-magnetic impurity
for this study. However, there is a seemingly discrepancy between
our early data in LaFeAsO$_{0.9}$F$_{0.1}$, where $T_c$ is
essentially unaffected by Zn-impurity\cite{LaOZnAs}, and a
following report showed a severe suppression of $T_c$ due to
Zn-impurity in the oxygen-deficient LaFeAsO$_{1-\delta}$
samples\cite{nims}. In this Letter, we report a systematic study
of the effect of Zn-impurity on superconductivity in
LaFeAsO$_{1-x}$F$_x$ in different doping regimes. We have observed
strong doping dependence of the Zn-impurity effect on $T_c$. $T_c$
is enhanced in the underdoped regime ($x=0.05$), remains
essentially unchanged at the optimal doping ($x=0.10$), and is
severely suppressed at the overdoped regime ($x=0.15$). Our
results suggest a switch of the symmetry of the SC order
parameters from a $s$-wave to $s_{\pm}$ or $d$-wave states as the
charge carrier increases in FeAs-based superconductors. The
increase in $T_c$ at low electron doping may be explained as a
result of the suppression of the magnetism upon Zn-doping, which
is in favor of the superconductivity.

\section{Experimental}

Polycrystalline samples of LaFe$_{1-y}$Zn$_y$AsO$_{1-x}$F$_x$ ($x$
= 0.05, $y$ = 0, 0.02, 0.04, 0.06; $x$ = 0.15, $y$ = 0, 0.02) were
synthesized by solid state reaction method. Details on the sample
preparation can be found in the previous report\cite{LaOZnAs}. The
phase purity of the samples was investigated by powder X-ray
diffraction (XRD) using a D/Max-rA diffractometer with
Cu-K$_{\alpha}$ radiation and a graphite monochromator. Lattice
parameters were calculated by a least-squares fit using at least
20 XRD peaks. Chemical analysis by energy-dispersive x-ray (EDX)
microanalysis was performed on an EDAX GENESIS 4000 x-ray analysis
system affiliated to a scanning electron microscope (SEM, model
SIRION).

The electrical resistivity was measured on bar-shaped samples
using a standard four-probe method. The measurements of resistance
and Hall effect were performed on a Quantum Design Physical
Property Measurement System (PPMS-9). DC magnetization were
measured on a Quantum Design Magnetic Property Measurement System
(MPMS-5).

\section{Results and discussion}

Figure 1 shows characterization of
LaFe$_{1-y}$Zn$_y$AsO$_{1-x}$F$_x$ samples by the XRD experiments.
The inset shows the variations of lattice constants with Zn
content for both F-underdoped and F-overdoped samples. All the XRD
peaks can be well indexed with the tetragonal ZrCuSiAs-type
structure, indicating that all the samples are single phase
without foreign phases. The $a$-axis shrinks slightly with the Zn
doping, while the $c$-axis increases, similar to the previous
report\cite{LaOZnAs}. The EDX measurements found that the actual
Zn content ($x$) is very close to the nominal composition, and the
variation of Zn content in the samples is less than 5\%. These
results indicate that Zn has successfully substituted Fe.

Figure 2 shows the temperature dependence of resistivity ($\rho$)
for LaFe$_{1-y}$Zn$_y$AsO$_{1-x}$F$_x$ from 2 K to 300 K and the
inset shows the temperature dependence of DC susceptibility for
the same samples. For the zinc-free LaFeAsO$_{0.95}$F$_{0.05}$
sample, the resistivity shows a SC transition at 16.8 K (defined
as the midpoint), consistent with the previous
report\cite{LaFFeAs}. In the underdoped regime, as shown in
Fig.2(a), with the increase of Zn content, to our surprise, $T_c$
increases to 19.2 K and 22.7 K for $y$ = 0.02 and 0.06
respectively. The $T_c$ values determined from the magnetic
susceptibility exhibits consistent variation with resistivity
data. The estimated Meissner volume fraction indicates bulk nature
for the superconductivity. Obviously, the above result is in
contrast to the case of Zn-doped high-$T_c$ cuprates\cite{Xiao}
where $T_c$ always decreases sharply with Zn doping.

The resistivity and DC susceptibility of overdoped
LaFe$_{1-y}$Zn$_y$AsO$_{0.85}$F$_{0.15}$ ($y$ = 0 and 0.02)
samples were shown in Fig. 2 (b). Without Zn doping, $T_c$
(midpoint) is about 9.6 K. With only 2\% Zn doping, $T_c$ is
severely suppressed to below 2 K. We did not observe the
diamagnetism in susceptibility in this sample, consistent with the
resistivity measurement. These results are in contrast to the case
of underdoped LaFe$_{1-y}$Zn$_y$AsO$_{0.95}$F$_{0.05}$ sample and
indicate that Zn doping in overdoped region does sharply suppress
superconductivity. A similar sharp suppression of $T_c$ with Zn
doping was also reported in the oxygen-deficient
LaFeAsO$_{1-\delta}$ where the sample is apparently overdoped
according to the oxygen content\cite{nims}.

Fig. 3 shows temperature dependence of Hall coefficient $R_{H}$
for LaFe$_{1-y}$Zn$_y$AsO$_{1-x}$F$_x$ samples. For all the
samples, the negative $R_{H}$ signal indicates that electron-type
charge carrier is dominant in the whole temperature region. For
the underdoped LaFe$_{1-y}$Zn$_y$AsO$_{0.95}$F$_{0.05}$ samples,
$R_{H}$ of the Zn-free sample exhibits a strong $T$-dependence and
drops sharply below 100 K, which may be attributed to the residual
spin-density-wave (SDW) order or fluctuations. $R_H$ finally goes
to zero as the samples enter into superconducting state. In the
parent compound LaFeAsO, $R_H$ drops sharply due to opening of SDW
gap below the structural/magnetic transition temperature
\cite{McGuire}. With increasing Zn content, the sharp drop in
$R_H$ is gradually suppressed, indicating that the residual SDW
order or fluctuations are further suppressed by Zn impurities.
Such an effect of Zn impurities on the SDW order has also been
observed in the parent compound LaFeAsO \cite{LaOZnAs}. Meanwhile,
the room-temperature $R_{H}$ remains almost unchanged with Zn
doping, indicating that Zn doping does not change the charge
carrier density. This is consistent with the band calculation
result which predicts that Zn 3$d$ electrons are deep below the
Fermi level\cite{Singh2009}. For the overdoped
LaFe$_{1-y}$Zn$_y$AsO$_{0.85}$F$_{0.15}$ samples [Fig.3 (b)],
$R_{H}$ of both the Zn-free and 2\% Zn-doped samples exhibit very
weak $T$-dependence and again the change in the room-temperature
$R_H$ is very small, indicating that the charge carrier density
essentially does not change with Zn doping. It should be noted
that $R_H$ drops quickly due to SC transition for the Zn-free
sample while it remains constant for Zn-doped sample as $T$ goes
to zero.

The effect of Zn-impurity in the different doping levels is
summarized in a phase diagram of $T_c$ versus Zn content, as shown
in Fig. 4. In the underdoped regime, i.e. 5\% F doped samples,
$T_c$ remarkably increases with increasing Zn-doped concentration,
while in the nearly optimally-doped regime,
LaFe$_{1-y}$Zn$_y$AsO$_{0.9}$F$_{0.1}$, $T_c$ almost does not
change with Zn doping, or it is even enhanced slightly at low Zn
content. In the overdoped regime, i.e.,
LaFe$_{1-y}$Zn$_y$AsO$_{0.85}$F$_{0.15}$, $T_c$ sharply decreases
to below 2 K even with only 2\% Zn doping.

Recently the Zn doping effect has also been studied in the
hole-type pnictide superconductor (Ba,K)Fe$_2$As$_2$ \cite{WenHH}.
Wen and his collaborators\cite{WenHH} confirmed that
superconductivity is robust against the non-magnetic Zn doping in
the nearly optimally-doped Ba$_{0.6}$K$_{0.4}$Fe$_2$As$_2$,
consistent with our result in the optimally doped regime.
Meanwhile, a severe suppression of $T_c$ was reported in a
oxygen-deficient 1111 phase LaFe$_{1-x}$Zn$_x$AsO$_{0.85}$ which
is apparently overdoped according to the oxygen content
\cite{nims}, in agreement with our result in the overdoped regime.
It should also be noted that the partial substitution of other
transition metal elements such as Co, Ni, Ru, Rh, Pd, and Ir, not
only have negligible pair-breaking effect, but even can induce
superconductivity. The 3$d$ electrons of these ions (such as Co,
Ni, and Ru) are believed to be itinerant according to the first
principle calculations \cite{FangZ}. In contrast, the 3$d$
electrons of Zn ions are localized without any local moments
\cite{Singh2009}.

Since the impurity effect is very sensitive to the symmetry of the
SC order parameter, the contrastive difference in the Zn impurity
effect between the underdoped and overdoped regimes cannot easily
be explained within the same pairing symmetry state.
 Instead, our result strongly suggests a switch of the pairing
symmetry from an impurity-insensitive pairing state to an
impurity-sensitive pairing state. Since other experimental results
such as the angle-resolved photoemission spectroscopy (ARPES)
measurements \cite{ARPES} have found that the underdoped or
optimally doped FeAs-based superconductors should have full
superconducting gaps, our present data suggests a $s$-wave pairing
in the underdoped and optimally doped regime. Meanwhile $s_{\pm}$
or $d$-wave pairings are strongly suggested in the overdoped
regime. Iron-based SC compounds have hole Fermi pockets centered
at the $\Gamma$ point in the Brillouin zone and electron pockets
centered at the zone corner (M point). The $s$-wave and
$s_{\pm}$-wave states correspond to the same and opposite signs of
the relative order parameters between the hole and electron Fermi
pockets, which is determined by the sign of the Josephson coupling
between the Fermi pockets. A switch from the $s$-wave to $s_{\pm}$
wave states is in accordance to the sign change of the Josephson
coupling from attractive at the underdoped to repulsive at
over-doped regimes.

Below we shall further elaborate the scenario of a switch between
pairing states. The impurity effect in the $s_{\pm}$-wave state
was studied by Bang et al.\cite{Bang} and by Onari and
Kontani\cite{Kontani}.
 Theoretical calculations
have shown that in the strong scattering limit the non-magnetic
impurity effect to $s_{\pm}$-wave state is severe and similar to
the effect on $d$-wave SC state\cite{Bang}. Their theory should
apply to Zn-impurity located in the Fe-planes, where the
scattering is strong. The severe suppression of $T_c$ in the
overdoped regime may thus be well explained within the scenario of
$s_{\pm}$-wave symmetry. Moreover, the $d$-wave pairing is
unlikely according to the ARPES studies in the overdoped regimes
\cite{DingH,DingH1}. On the other hand, the insensitivity of the
impurity effect in the under-doped and optimally doped regimes is
not compatible with the $s_{\pm}$ pairing. Note that the
suppression of $T_c$ by impurities in the $s_{\pm}$ wave is not
dependent on the charge carrier doping level. Our data strongly
suggest that in the underdoped and optimally doped regime, the SC
pairing is likely $s$-wave, essentially unaffected by the
Zn-impurity according to Anderson theorem\cite{Anderson}. The
enhancement of $T_c$ with Zn-impurity in the underdoped regime may
be explained as the result of the suppression of AFM order, which
in turn enhances superconductivity.

We remark that the possible changes in pairing symmetry with
doping level were also proposed in a recent NMR study on the
P-doped BaFe$_2$As$_2$ system \cite{BaP122}, and in a theoretical
proposal for time reversal symmetry breaking state\cite{ZhangSC},
and in a theory for a possible switch between nodeless and nodal
pairings by changing the pnictogen height measured from the Fe
plane \cite{Switch}.

\section{Conclusion}

In summary, we have studied systematically the effect of
non-magnetic Zn-impurity on superconductivity in various doping
regimes of LaFeAsO$_{1-x}$F$_x$ systems. The Zn-impurities do not
suppress superconductivity in the underdoped and optimally doped
regimes, but severely suppress superconductivity in the overdoped
regime. Our results suggest a switch of the pairing symmetry from
Zn-impurity insensitive $s$-wave at underdoped and optimally doped
regime to impurity-sensitive pairing state (likely $s_{\pm}$-wave)
at overdoped regime. The enhancement in $T_c$ with increasing Zn
content at the low F-doping regime could result from the
suppression of AFM order or fluctuation in FeAs layers which
competes with superconductivity.

\section*{Acknowledgments}
This work is supported by the National Science Foundation of China
(Grant nos 10634030 and 10931160425), PCSIRT (IRT-0754), the
National Basic Research Program of China (Grant No. 2007CB925001).
We also acknowledge partial support from Hong Kong RGC grant HKU
7068/09P and NSF/RGC N-HKU 726/09.

\pagebreak[4]

\begin{figure}
\includegraphics[width=8cm]{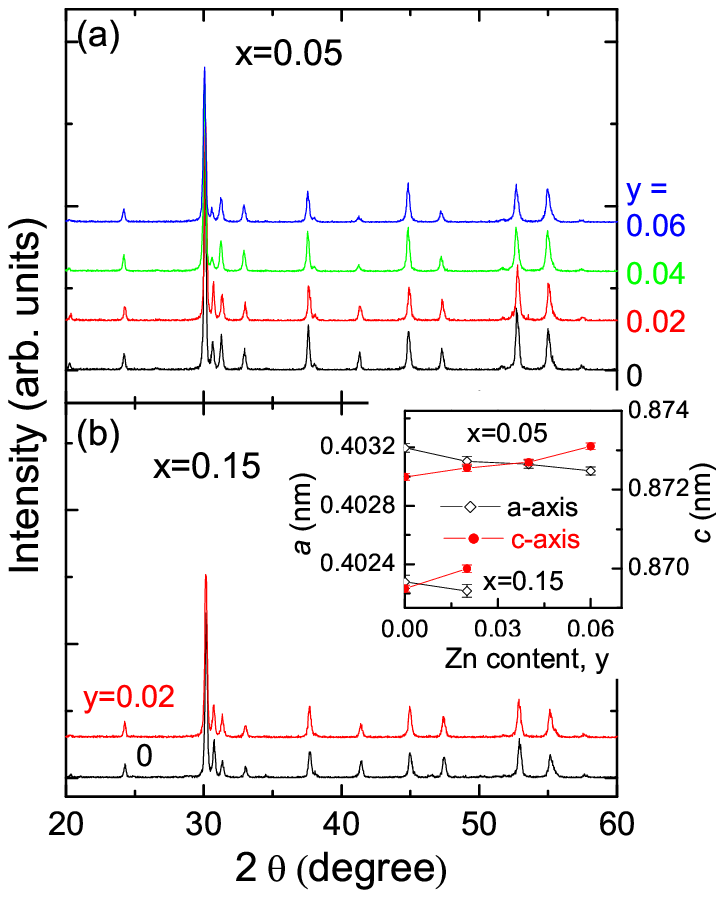}
\caption{\label{Fig.1} (color online). Structural characterization of
LaFe$_{1-y}$Zn$_y$AsO$_{1-x}$F$_x$ samples. (a) and (b) Powder
X-ray diffraction patterns for $x$=0.05 and 0.15, respectively.
Inset: Lattice parameters as functions of Zn content.}
\end{figure}

\begin{figure}
\includegraphics[width=8cm]{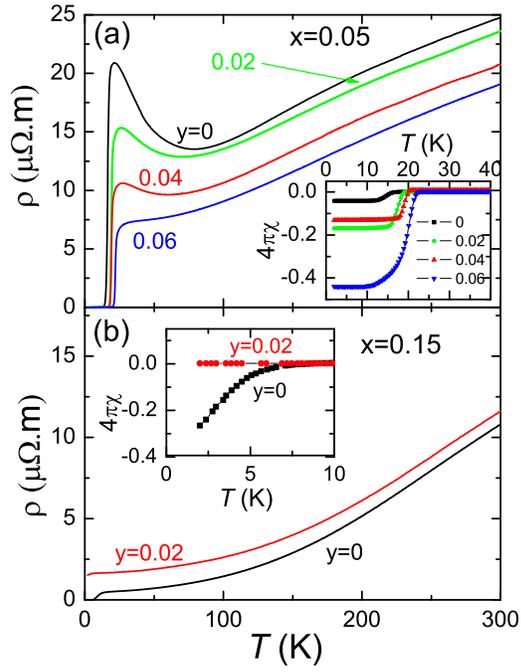}
\caption{\label{Fig.2} (color online). Temperature dependence of resistivity
($\rho$) for the LaFe$_{1-y}$Zn$_y$AsO$_{1-x}$F$_x$ samples. The
inset: diamagnetic transitions under a magnetic field of 10 Oe
with field-cooling (FC) mode. (a) $x$ = 0.05, $y$ = 0, 0.02, 0.04,
and 0.06 ; (b) $x$ = 0.15; $y$ = 0 and 0.02.}
\end{figure}

\begin{figure}
\includegraphics[width=8cm]{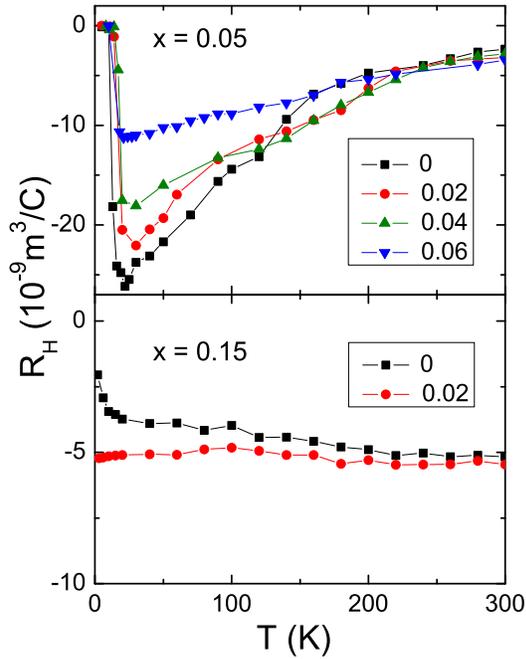}
\caption{\label{Fig.3}(color online) Temperature dependence of Hall coefficient
$R_{H}$ measured at 5 T for LaFe$_{1-y}$Zn$_y$AsO$_{1-x}$F$_x$.
(a) $x$ =0.05, $y$ = 0, 0.02, 0.04, 0.06; (b) $x$ = 0.15, $y$ = 0,
0.02.}
\end{figure}

\begin{figure}
\includegraphics[width=8cm]{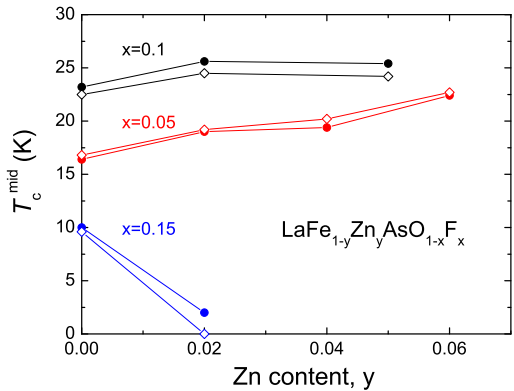}
\caption{\label{Fig.4}(color online) Superconducting transition temperatures
versus Zn content in LaFe$_{1-y}$Zn$_y$AsO$_{1-x}$F$_x$. Solid and
open symbols refer to $T_c$ determined from the measurements of
resistivity (midpoint) and susceptibility (onset point),
respectively. The data of $x$ = 0.1 were taken from the previous
report.\cite{LaOZnAs}}
\end{figure}

\end{document}